\documentclass[lettersize,journal]{IEEEtran}

\usepackage{amsmath,amsfonts}
\usepackage{algorithmic}
\usepackage{array}
\usepackage[caption=false,font=normalsize,labelfont=sf,textfont=sf]{subfig}
\usepackage{textcomp}
\usepackage{stfloats}
\usepackage{url}
\usepackage{verbatim}
\usepackage{graphicx}
\def\BibTeX{{\rm B\kern-.05em{\sc i\kern-.025em b}\kern-.08em
		T\kern-.1667em\lower.7ex\hbox{E}\kern-.125emX}}

\usepackage{colortbl}
\usepackage{bm}
\usepackage{fancyhdr}       

\usepackage{url}       

\usepackage{cite}

\usepackage{amssymb}

\usepackage{stfloats}
\usepackage{indentfirst}
\usepackage{cases}
\usepackage{algorithm}

\usepackage{epstopdf}
\usepackage{diagbox}
\usepackage{subfig}
\usepackage{caption}
\usepackage{tikz}
\usepackage{color}
\usepackage{bm}

\usepackage{algorithm}
\usepackage{colortbl}
\usepackage{bm}
\usepackage{epstopdf}

\usepackage{url}       
\usepackage{cite}
\usepackage{extarrows}

\usepackage{stfloats}

\usepackage{cases}
\usepackage{algorithm}
\usepackage{multirow}
\usepackage{colortbl}
\usepackage{bm}
\usepackage{epstopdf}

\usepackage{url}       

\usepackage{cite}
\usepackage{extarrows}

\usepackage{stfloats}

\usepackage{cases}
\usepackage{algorithm}
\usepackage{multirow}

\newcommand{\ls}[1]
    {\dimen0=\fontdimen6\the\font
     \lineskip=#1\dimen0
     \advance\lineskip.5\fontdimen5\the\font
     \advance\lineskip-\dimen0
     \lineskiplimit=.9\lineskip
     \baselineskip=\lineskip
     \advance\baselineskip\dimen0
     \normallineskip\lineskip
     \normallineskiplimit\lineskiplimit
     \normalbaselineskip\baselineskip
     \ignorespaces
    }

\graphicspath{{Visio-File-0731-final/}{Simulation-1213/}}

\begin{document}
\title{{Statistical QoS Provisioning Architecture for 6G Satellite-Terrestrial Integrated Networks}}

\author{Jingqing Wang, Wenchi Cheng, Wei Zhang, Hui Liang\vspace{-2pt}
\\
\thanks{\ls{.5}
\setlength{\parindent}{2em}

This work was supported by the Key Area Research and Development Program of Guangdong Province under Grant 2020B0101110003, in part by the National Natural Science Foundation of China under Grant 62341132, the National Key Research and Development Program of China under Grant 2021YFC3002102, the Key Research and Development Plan of Shaanxi Province under Grant 2022ZDLGY05-09, and the Natural Science Basic Research Program of Shaanxi under Grant 2024JC-YBQN-0642.

Jingqing Wang and Wenchi Cheng are with Xidian University, Xi'an, 710071, China (e-mails: jqwangxd@xidian.edu.cn; wccheng@xidian.edu.cn).

Wei Zhang is with the School of Electrical Engineering and Telecommunications, The University of New South Wales, Sydney, Australia (e-mail: w.zhang@unsw.edu.au).
}
}


\maketitle



\begin{abstract}
The emergence of massive ultra-reliable and low latency communications (mURLLC) as a category of age/time/reliability-sensitive service over 6G wireless networks has received considerable research attention, which has presented unprecedented challenges.
As one of the key enablers for 6G, satellite-terrestrial integrated networks have been developed to offer more expansive connectivity and comprehensive 3D coverage in space-aerial-terrestrial domains for supporting 6G mission-critical mURLLC applications while fulfilling diverse and rigorous quality-of-service (QoS) requirements.
In the context of these mURLLC-driven satellite services, data freshness assumes paramount importance, as outdated data may engender unpredictable or catastrophic outcomes.
To effectively measure data freshness in satellite-terrestrial integrated communications, age of information (AoI) has recently surfaced as the new dimension of QoS metrics to support time-sensitive applications.
It is crucial to design new analytical models that ensure stringent and diverse QoS metrics bounded by different key parameters, including AoI, delay, and reliability, over 6G satellite-terrestrial integrated networks.
However, due to the complicated and dynamic nature of satellite-terrestrial integrated network environments, the research on efficiently defining the new statistical QoS provisioning schemes while taking into account varying degrees of freedom has still been in their infancy.
To remedy these deficiencies, in this paper we develop statistical QoS provisioning schemes over 6G satellite-terrestrial integrated networks in the finite blocklength regime.
Particularly, we firstly introduce and review key technologies for supporting mURLLC.
Secondly, we formulate a number of novel fundamental statistical-QoS metrics in the finite blocklength regime.
Finally, we conduct a set of simulations to validate and evaluate our developed statistical QoS provisioning schemes over satellite-terrestrial integrated networks.
\end{abstract}


	\section{Introduction}\label{sec:intro}

With 5G wireless networks being deployed globally, researchers are actively conceptualizing future 6G mobile wireless networks\textsuperscript{\cite{9861699}} for supporting overwhelmingly unparalleled modern wireless applications with increasingly rigorous and diverse {technical performance requirements, such as stringent end-to-end delay, ultra-high reliability, unparalleled data rate, among others.}
One of the key focuses of 6G development is the design of systems to enable \textit{massive Ultra-Reliable Low-Latency Communications} (mURLLC)\textsuperscript{\cite{9311792}}, requiring {diverse quality-of-services (QoS)} guarantees to ensure their timely and reliable delivery.
However, the crucial obstacle presented by expansive connectivity and comprehensive coverage imposes a pivotal challenge in supporting 6G mission-critical mURLLC applications.
To enable comprehensive 3D coverage for various 6G services using space-aerial-terrestrial integration, satellite communication systems have been developed as a potential alternative for terrestrial communication systems to offer global coverage and seamless connectivity, while guaranteeing stringent QoS benchmarks requisite.
On the other hand, conventional terrestrial networks still play an important role in providing low-cost and high-speed wireless services considering densely populated areas.
Therefore, integrating satellite and terrestrial networks can leverage the advantages of both systems, enabling ubiquitous network service.

{Since} end-to-end satellite communication can effectively provide seamless coverage and lower latency in comparison to terrestrial networks, particularly in remote rural areas and isolated deserts, satellite-terrestrial integrated networks present significant potential in facilitating various emerging time-sensitive applications, such as intelligent traffic, environment monitoring, and disaster relief.
{In the context of these real-time applications, ensuring a variety of stringent QoS benchmarks becomes of paramount importance, as outdated and unreliable data may engender unpredictable or catastrophic outcomes.
Nevertheless, {the complicated and dynamically evolving nature of satellite-terrestrial integrated network environments inevitably engenders system performance requirements,} thus, in turn, would significantly impact QoS assurances in terms of data freshness, delay, and reliability.	
In light of the aforementioned challenges, it is imperative to develop potential diverse QoS measurement and control strategies, guaranteeing the stringent mURLLC requirements over 6G satellite-terrestrial integrated networks.}

{As a result, providing distinct levels of QoS measurement with specific data freshness/delay/reliability constraints for diverse categories of age/time/error-sensitive wireless multimedia data constitutes a pivotal element within the framework of 6G satellite-terrestrial integrated networks.
Traditionally, researchers have established} delay-bound QoS provisioning for supporting real-time wireless services.
The inherently unstable nature of wireless fading channels and the intricate, diverse, and ever-changing structures of 6G satellite-terrestrial integrated networks make it challenging to uphold the conventional deterministic QoS criteria for upcoming {age/time/error-sensitive} multimedia data transmission.
Towards this end, the \textit{statistical QoS provisioning theory} has been developed as a potent strategy for defining and implementing QoS guarantees.
{The authors of Ref. {\cite{9834874} have defined and characterized the novel statistical QoS metrics in terms of delay and error-bounded QoS exponents.}

Nevertheless, the ever-increasing amount of {age, delay, and error sensitive} multimedia traffic necessitates 6G networks to meet increasingly complex and demanding QoS requirements.
More specifically, in scenarios involving space communication, it becomes essential to effectively measure and manage the data freshness for guaranteeing mURLLC services.
To effectively measure/control data freshness, \textit{age of information} (AoI)\textsuperscript{\cite{Age2022}-\cite{9109636}} has recently surfaced as a novel QoS metric { that quantifies the freshness of information at the receiver's end for accurately reflecting data packets' update speed in short-pack-based status update systems, particularly for satellite applications\textsuperscript{\cite{9834874}}.
It represents the time elapsed since the most recently received update was generated at the source and provides insights into how ``up-to-date" the information, irrespective of the transmission delays incurred along the communication path.}	
Due to the limited number of information bits in satellite status updates and the need for prompt delivery to remote destinations, long codewords impose significant demands on receiver's storage and computational resources, thereby diminishing the system's timeliness.
Thus, small-packet communications, such as finite blocklength coding (FBC)\textsuperscript{\cite{Poor2023}-\cite{2023performance}}, play a crucial role in measuring/controlling the AoI metric across satellite-terrestrial integrated networks.

Despite the diligent efforts from both academia and industry for guaranteeing {stringent} QoS requirements of mURLLC services via satellite-terrestrial integrated networks, most previous studies have focused on investigating QoS metrics in terms of the delay-bound violating probability, without considering the non-vanishing decoding error probability {and the AoI violation probability}.
It is crucial to design and model satellite-terrestrial integrated wireless network architectures while taking into account { stringent and diverse age/delay/error-rate bounded QoS guarantees by defining more fundamental statistical QoS performance metrics and characterizing their analytical relationships, such as delay-bound-violation probability, decoding error probability, and AoI violation probability, in the finite blocklength regime.}
However, due to the lack of comprehensive grasp of QoS-driven fundamental performance-limits and modeling analyses over satellite-terrestrial integrated networks, how to efficiently identify and define novel FBC-based QoS metrics bounded by \textit{AoI}, \textit{delay}, and \textit{error-rate} as well as the associated relationships, including diverse \textit{QoS-exponent functions}, is still a challenging task.

To address these challenges, in this paper we develop a series of fundamental statistical QoS metrics and their modeling/controlling techniques for mURLLC over 6G satellite-terrestrial integrated networks.
Firstly, for supporting mURLLC, we introduce and review key techniques and potential solutions within the purview of standards development entities.
Secondly, we develop a set of new statistical QoS metrics and controlling functions.
Finally, a set of simulations is performed to verify, assess, and examine the effectiveness of the proposed statistical QoS provisioning schemes over satellite-terrestrial integrated 6G wireless networks.

The rest of this paper is structured as follows. Section~\ref{sec:sys} reviews the most recent advancements and potential solutions.
Section~\ref{sec:EC1} develops a set of new and fundamental statistical QoS metrics.
Section~\ref{sec:results} validates and evaluates our developed modeling techniques and controlling schemes. The paper concludes with Section~\ref{sec:conclusion}.

\section{{QoS Provisioning for 6G Satellite-Terrestrial Integrated Networks: Challenges and Open Issues}}\label{sec:sys}

\begin{figure*}[!t]
	\centering
	\captionsetup{labelformat=default,labelsep=space} 
	\includegraphics[scale=0.465]{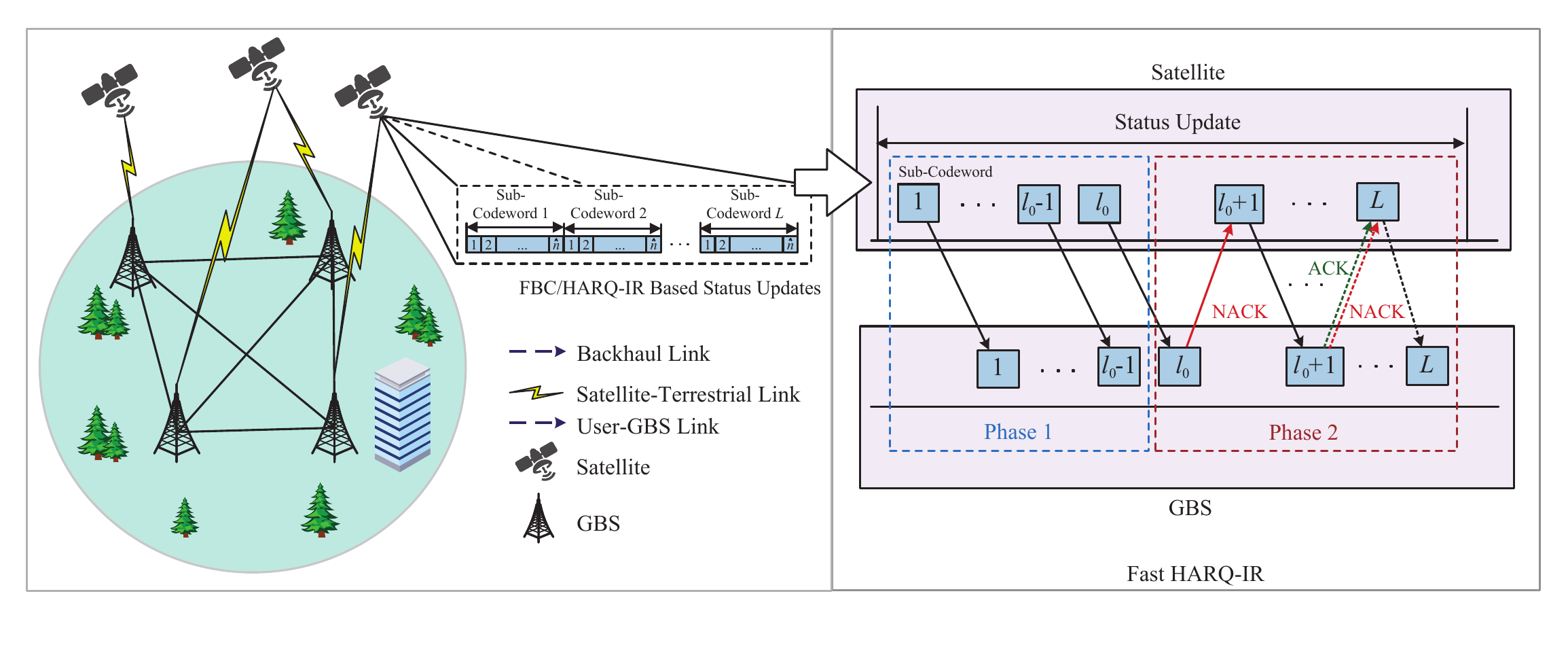}
	\caption{The system model for the satellite-terrestrial integrated wireless networks using fast HARQ protocol in the finite blocklength regime}
	\label{fig:1}
\end{figure*}

Consider the scenario of delivering downlink Telemetry (TM) status updates over satellite-terrestrial integrated wireless networks.
Fig.~\ref{fig:1} shows satellite-terrestrial integrated system architecture model for remote TM status update transmissions, which consists of $K_\text{G}$ ground base stations (GBSs).
Specifically, we focus on the satellite's responsibility to monitor dynamic physical processes by transmitting remote status updates.
The satellites capture these status messages as needed and subsequently transmits the updates to GBSs while guaranteeing stringent and diverse QoS constraints for mURLLC.
The distinctive characteristics of satellite communications introduce many new challenges when attempting to adapt terrestrial 5G NR technologies for these contexts.

{Researchers have addressed the incorporation of QoS provisioning over satellite-terrestrial integrated architecture through analyzing relevant QoS mechanisms in different layers and evaluating the corresponding influence. 
However, there are still numerous unresolved issues crucial for ensuring QoS guarantees over satellite-terrestrial integrated wireless networks. 
These encompassed concerns related to bandwidth allocation and enhancements of the transmitting protocols tailored for satellite communications.
Due to space limitations, we focus on part of the most critical aspects for the satellite-terrestrial integrated architecture when taking into account the large delays and reliability issues, many other factors will need a proper analysis and characterisation to fully implement 6G satellite-terrestrial integrated networks, as, for instance, the downlink status update transmissions in LEO scenarios.}

{\subsection{The Preliminary Analyses for Delay-Bounded QoS in Satellite-Terrestrial Integrated Wireless Networks}}
	
The implementation of statistical delay-bounded QoS guarantees has allowed for the investigation of queuing behaviors that are influenced by the fluctuations of both arrival and service processes over time.
To meet statistical tail-latency requirements, we present the delay-bounded QoS provisioning, which characterizes the probability of processing queue length surpassing a specified threshold.

\textit{Definition 1: The delay-bounded delay-QoS exponent:} By using the \textit{large deviations principle} (LDP), {under sufficient conditions, the queueing process approaches a random variable $Q(\infty)$ in distribution such that
	\begin{equation}\label{equation1}
		-\lim_{Q_{\text{th}}\rightarrow\infty}\frac{\log\left(\text{Pr}
			\left\{Q(\infty)>Q_{\text{th}}\right\}\right)}{Q_{\text{th}}}=\theta_{\text{delay}}
	\end{equation}
	where $Q_{\text{th}}$ represents the overflow threshold for the queueing systems and $\theta_{\text{delay}}$ $(\theta_{\text{delay}}>0)$ is defined as the \textit{delay-QoS exponent} of queuing delay, which measures the exponential decay rate of the delay-bounded QoS violation probabilities.}

The delay-bounded QoS exponent serves as an indicator of the level of stringency with bounded delay as the threshold advances.
Specifically, a sufficiently small delay-bounded QoS exponent enables the satellite-terrestrial integrated system to accommodate an arbitrarily prolonged delay, whereas a sufficiently large delay-bounded QoS exponent renders the system intolerant of any delay.
{For a given designated delay threshold, the delay violation probability characterizes the tail behaviors of the queueing delay, which is a crucial factor in determining the fundamental performance-limits of statistical delay-bounded QoS.

However, the existing research works mainly focus on characterizing system performances without considering stringent end-to-end delay and reliability requirements for mURLLC. 
It is crucial to design and model satellite-terrestrial integrated wireless network architectures while taking into account diverse QoS requirements by identifying and defining more fundamental statistical QoS performance metrics and characterizing their analytical relationships, such as various QoS-exponent functions, in the finite blocklength regime. 
Moreover, many existing publications delve into the typical challenges that arise during the implementation of satellite networks, regardless of their specific integration with terrestrial infrastructures.
In our examination of the specific satellite-terrestrial integrated scenario, the investigations of end-to-end QoS provisionings with statistical delay and error-rate constraints, are still in their infancy. 
This can be attributed to both insufficient research efforts on the mURLLC-oriented 6G network protocol designs and lack of the comprehensive understanding on various QoS-driven fundamental performance-limits and their modeling analyses.}

{\subsection{Protocol Adaptations for Satellite-Terrestrial Integrated Wireless Networks}

Researchers have investigated the modification of protocols to enhance the suitability and operational efficacy for satellite communications.}
The HARQ protocols have been developed to allow receivers to exploit small data packets from previous HARQ transmissions to improve the probability of successful decoding for a specific data packet.
{As shown in Fig.~\ref{fig:1},} employing the FBC technique enables encoding the status update packet into a small codeword consisting of $n$ channel uses for reducing access latency and decoding complexity.
Each finite-blocklength codeword with length $n$ is divided into $L$ sub-codewords of $\widehat{n}$ symbols, and the sub-codewords are transmitted successively during consecutive time slots.
The system capacity relies on various factors, such as the blocklength $\widehat{n}$, retransmission round $l$, and transmitting packet size.
Note that when the number of HARQ retransmissions increases, the amount of redundancy is introduced, leading to a decreased system capacity.
This decrease in coding rate results in a corresponding reduction in the decoding error portability, thereby enhancing the reliability of wireless communication systems.
{Nevertheless, in scenarios characterized by extended round-trip times (RTTs) and the stop-and-wait protocol, communication throughput may be compromised as transmissions will stall when all HARQ processes await feedback.
Accordingly,} HARQ protocols can introduce significant delays due to the need for retransmissions and multiple decoding attempts, resulting in the degradation of QoS provisioning.

To mitigate these challenges stemming from extended delays, it becomes essential to explore avenues for enhancing the management of HARQ protocol.
Therefore, we need to carefully design the proposed satellite-terrestrial integrated schemes to ensure overall age, delay, and reliability based QoS requirements.
A common solution to such an issue involves the provision of support for a substantial number of parallel HARQ processes in order to be compatible and to offer the desired peak data rate over satellite link.
While awaiting an ACK/NACK, multiple HARQ processes operate concurrently to enhance throughput by enabling other processes to engage in simultaneous work on other packets while a process remains in waiting state for an ACK/NACK. 
{HARQ protocol plays a pivotal role in determining the number of parallel HARQ processes and required retransmissions, thereby influencing the end-to-end delay, buffer size, and payload capacity of the satellite. 
The research on comprehensively analyzing the delay components contained in the end-to-end delay while implementing HARQ has still been in their infancy.
Thus, simply expanding the number of parallel HARQ processes is considered undesirable.}

Towards this end, 3GPP has embarked on research efforts focusing on the above-mentioned concerns, while further endeavors are required to enhance the initial conceptual solution.
Potential designs including dynamic HARQ deactivation, ACK-free HARQ, and adaptive HARQ feedback mechanisms should be contemplated. 
{Strategies for judicious reuse of the same HARQ process prior to the completion of a full RTT should also be employed to prevent stalling. 
Moreover, examining the deactivation of HARQ feedback represents an additional facet to explore for a comprehensive understanding of potential advantages and drawbacks for the proposed satellite-terrestrial integrated schemes.}

{In particular,} researchers have proposed the fast HARQ protocol\textsuperscript{\cite{9686609}} as a potential design for guaranteeing statistical QoS, which eliminates some HARQ feedback signals and successive message decoding by estimating the number of HARQ retransmission rounds, as shown in Fig.~\ref{fig:1}.
In \textbf{Phase 1,} upon receiving the initial sub-codeword of a status update, the satellite proceeds with the estimation of the necessary number of HARQ retransmission rounds $l_{0}$ for ensuring successful decoding at the receiver.
The decoder remains inactive during $l_{0}$ transmission attempts.
Correspondingly, the transmitter determines the retransmission round and transmission rate.
In \textbf{Phase 2},  i.e., round $l = l_{0},\dots,L$, the satellite decodes the sub-codewords received and returns an ACK/NACK to the transmitter, thus saving decoding and feedback delays through the use of the fast HARQ protocol.

{\subsection{The End-to-End Delay-Bounded QoS Metrics Using HARQ}

	Due to the elevated altitudes involved, the extended end-to-end delay, including queuing delay, transmission delay, processing delay, and propagation delay, exerts a pronounced influence on various facets of the radio interface. 
	The end-to-end delay is intricately linked to both the one-time access delay and the HARQ retransmissions. 
	In light of statistical QoS theory, it is possible to approximately characterize the behavior of remote tails in relation to the delay violation probability while implementing the HARQ protocol.
	However, varying numbers of HARQ retransmissions contribute to distinct compositions of delay, affecting the queuing delay value, the count of processing delays, and propagation delays. 
}

\begin{figure}[!t]
	\captionsetup{labelformat=default,labelsep=space} 	\centering
	\includegraphics[scale=0.28]{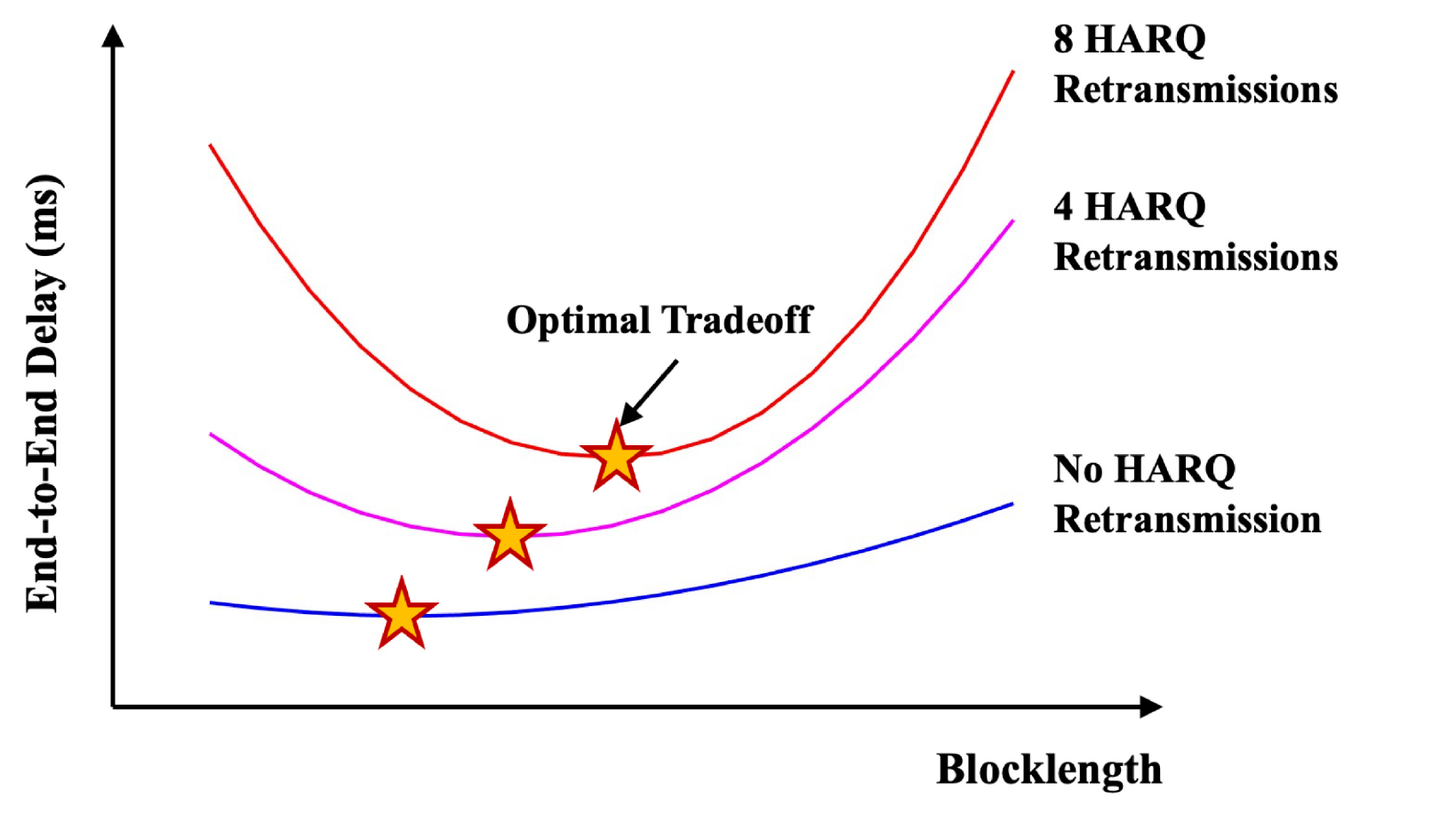}
	\caption{{The end-to-end delay versus the blocklength using HARQ in the finite blocklength regime.}}
	\label{fig:2}
\end{figure}

To support mURLLC services, {we need to investigate the end-to-end delay-bounded QoS metrics through using HARQ protocols and analyze the number of required HARQ retransmission rounds and their corresponding impact on statistical delay-bounded QoS provisioning schemes to improve the performance of satellite-terrestrial integrated wireless communication system.	
Specifically, the tradeoff relationship among queuing delay, transmission delay, and the number of HARQ retransmissions (including the propagation delays) within the intricate and dynamic contexts of satellite-terrestrial integrated wireless network environments need to be elucidated in the finite blocklength. 
This forms the basis for the development of the end-to-end delay-bounded QoS metrics.

Previous research has studied the tradeoff relationship between the queuing delay and transmission delay in the terrestrial system\textsuperscript{\cite{Adaptive2022}}.
Within the finite blocklength regime, reducing the blocklength in 5G NR leads to an increase in queuing delay and the number of retransmissions\textsuperscript{\cite{5GNR}}.
On the other hand, the transmission delay exhibits a gradual decrease as the blocklength diminishes. 
This implies that, as the blocklength decreases, there exist a tradeoff relationship among the transmission delay, queuing delay, and the number of HARQ retransmissions, as shown in Fig.~\ref{fig:2}. 
Optimal tradeoff is achieved when these three components reach an optimal balance, resulting in the minimum end-to-end delay. 
Fig.~\ref{fig:2} shows that when setting a large number of HARQ retransmissions $L$, there is an observable increase in the end-to-end delay. 
This observation suggests that the HARQ protocol has the potential to extend the propagation delay, necessitating the application of a fast HARQ protocol as a potential remedy.
However, within the intricate and dynamic contexts of satellite-terrestrial integrated wireless network environments, how to apply statistical QoS theory to characterize the relationship between the \textit{end-to-end delay-bounded QoS metrics} and the number of necessary HARQ retransmissions and find the optimal tradeoff point remain an open and challenging problem.}


\begin{figure}[!t]
	\captionsetup{labelformat=default,labelsep=space} 	\centering
	\includegraphics[scale=0.34]{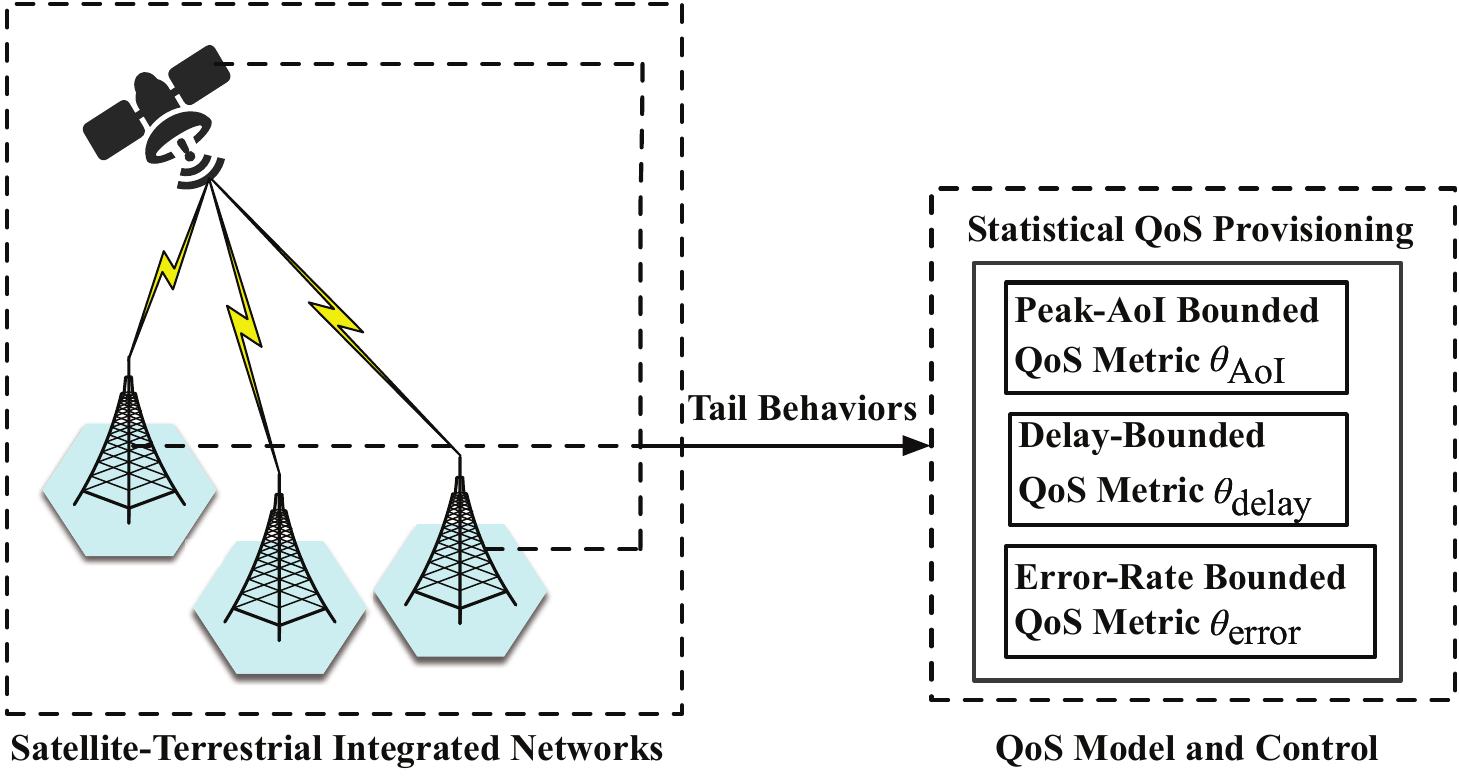}
	\caption{The statistical QoS provisioning metrics for the controlling functions over satellite-terrestrial integrated networks.}
	\label{fig:02}
\end{figure}

\section{{Statistical QoS Control Metrics} Over Satellite-Terrestrial Integrated Networks in the Finite Blocklength Regime}\label{sec:EC1}

{One critical aspect to consider for designing satellite-terrestrial integrated wireless services is the efficient guarantee of low-tail requirements for age/time/reliability-sensitive data transmissions, which play a crucial role in the fundamental performance analysis and provide practical engineering guidance for designing, analyzing, and evaluating diverse statistical QoS provisioning schemes.
As illustrated in Fig.~\ref{fig:02}, to measure, analyze, and effectively manage the stochastic behaviors during the data transmissions via satellite, our investigation is focused on analyzing the essential analytical modeling metrics, especially the tail behaviors of the delay-bounded bounded QoS metric, peak-AoI bounded QoS metric, and the error-rate bounded QoS metric.
Diverse statistical QoS metrics can provide valuable insights into the worst-case scenario performances for our proposed schemes.
	However, the analysis of tail behaviors, i.e., the peak AoI violation probability, presents several challenges.
	For instance, meeting the demanding requirements for mURLLC services, such as stringent data freshness, latency, and reliability, becomes increasingly challenging as complexity increases.
	Moreover, suboptimal tail behaviors can have a negative impact on user experience, particularly as degrees of parallelism intensify. 
	As a result, our proposed modeling schemes prioritize facilitating QoS provisioning for low-tail cases. 
	This emphasis is directed towards evaluating the violation probability, ensuring efficient support for the specific requirements associated with low-tail considerations for age-sensitive data transmissions, rather than solely assessing average performance metrics.
	However, traditional statistical QoS theory only investigates low-tail metrics in terms of queueing delay, without analyzing tail behaviors for AoI and error-rate. The reduction of violation probability that AoI of status updates exceeds a given age constraint is of great significance for guaranteeing the data freshness in our proposed systems.}	
Consequently, in this section, we develop a set of new fundamental statistical-QoS controlling metrics {
to characterize the violation probabilities with regard to peak AoI through using HARQ-IR, which can severely impact network performance for the developed satellite-terrestrial integrated schemes using FBC.}

{Statistical delay-bounded QoS provisioning} has been proposed {in the previous section as an effective approach to ensure the probability guarantee for violating service constraints rather than average performance metrics.}
{However, traditional statistical QoS theory only investigates low-tail metrics in terms of queueing delay, without analyzing tail behaviors for AoI and error-rate.
Towards this end, to statistically measure/control delay, peak AoI, and error-rate associated with mURLLC, we develop the statistical delay, peak AoI, and error-rate bounded QoS provisioning framework, as depicted in Fig.~\ref{fig:3}.
The probabilities of such} violating service constraints are considered to be essential QoS metrics, which play a crucial role in the fundamental performance analysis and provide practical engineering guidance for designing, analyzing, and evaluating {novel statistical QoS provisioning schemes.}

\begin{figure*}[!t]
	\captionsetup{labelformat=default,labelsep=space} 	\centering
	\includegraphics[scale=0.34]{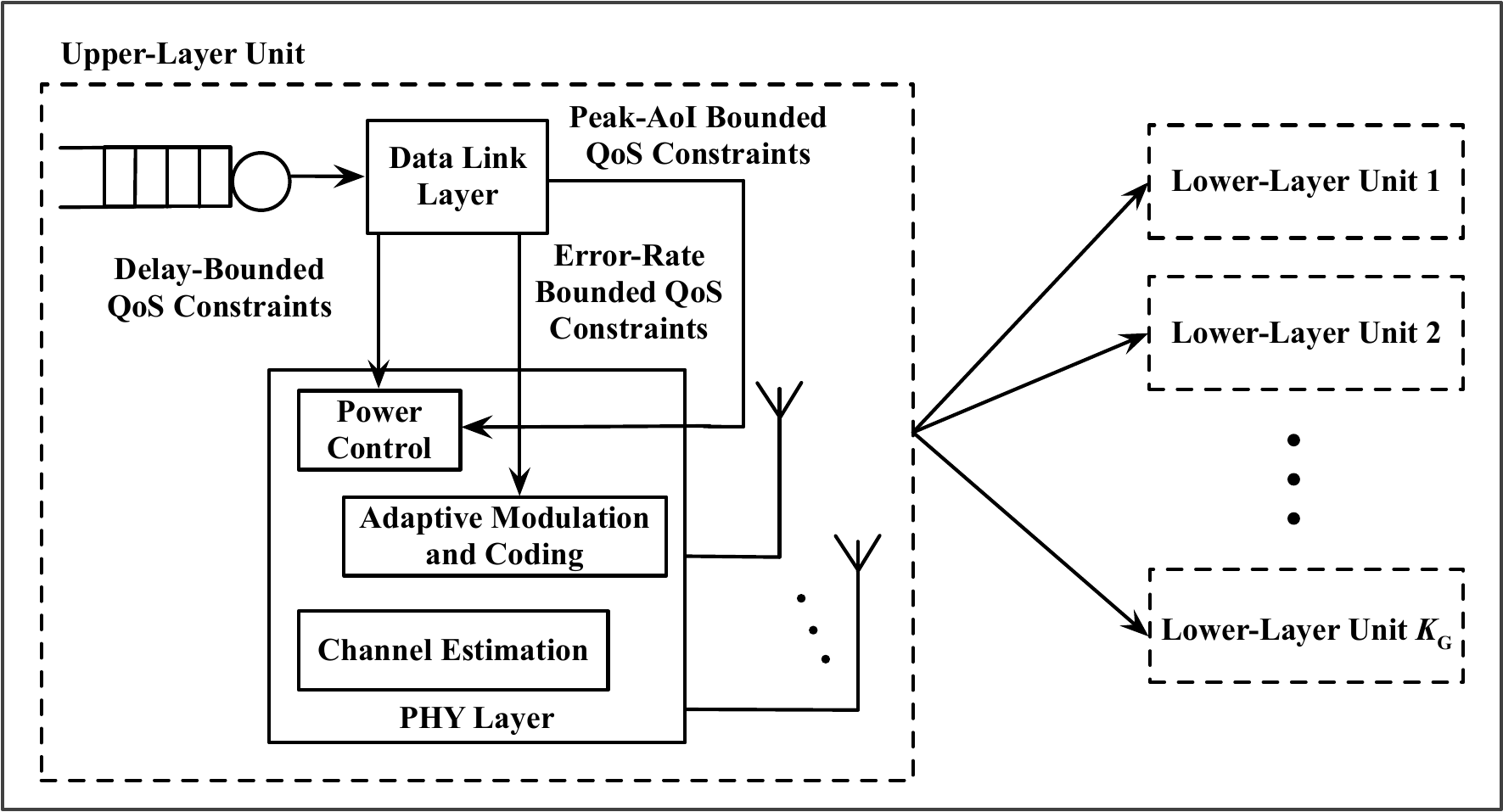}
	\caption{The framework of statistical diverse QoS provisioning for downlink satellite-terrestrial integrated networks.}
	\label{fig:3}
\end{figure*}

\subsection{The Peak-AoI Bounded QoS Metrics Using HARQ}\label{sec:pAoI}

\subsubsection{The AoI Metric Modeling in the Finite Blocklength Regime}

The TM status updates are transmitted from the satellite to GBSs via a wireless link susceptible to errors, with the objective of facilitating timely and reliable data processing and analysis.
To quantify and manage the timeliness of {status update packets,} we employ the AoI metric as a {new} fundamental QoS measure to indicate the recency of received data.
The service time for each {TM status update consists of the queuing delay, transmission delay, and propagation delays by implementing HARQ.} 
The transmission {delay is decided by the coding redundancy, while the propagation delay is dependent on the HARQ retransmission rounds and satellite} communication distance.
The AoI metric increases linearly in the absence of any new arrival updates, and upon receiving a successful update, it is reset to a lower value.

The total AoI over a period indicates the average delay of the updated data, representing the freshness of the sensory information.
By regarding the AoI as the crucial QoS metric, the challenge of pursuing obscured data freshness in satellite communications can be approached as a mathematical problem that can be solved through the implementation of optimization methods.
However, previous research mainly focuses on analyzing the average AoI to guarantee data freshness.
The average AoI provides a limited understanding of the latency by only characterizing the typical average latency of the satellite communication system, without considering the system's worst-case performance, which essentially serves as the primary impediment in ensuring mURLLC services.
Towards this end, we proposed to investigate the peak AoI as a potential solution to address the bottleneck issues in time-sensitive applications while using HARQ.

\subsubsection{The Peak-AoI Bounded QoS Exponent}
To support the peak-AoI bounded QoS provisioning, we propose to conduct a characterization of the probability of peak AoI violation, which can be determined by the probability that the peak AoI exceeds the designated peak AoI threshold.
{Accordingly, the peak AoI violation probability, denoted as $p^{\text{AoI}}$, can be determined by the probability that the peak AoI exceeds the designated peak AoI threshold, i.e.,
\begin{align}\label{equation49}
	p^{\text{AoI}}\triangleq\text{Pr}\left\{ P^{\text{AoI}}>\frac{A_{\text{th}}}{n}\right\}
\end{align}
where $A_{\text{th}}$ is the peak AoI violation threshold in channel uses through using FBC.
Then, we proceed to characterize the peak AoI violation probability through using the peak-AoI bounded QoS exponent, which is defined as follows.}

\textit{Definition 2: The peak-AoI bounded QoS exponent:} {By using the LDP, under sufficient conditions, the probability that peak AoI violation probability $P_{\text{AoI}}$ exceeds a certain threshold $A_{\text{th}}$ decays exponentially fast as the threshold $A_{\text{th}}$ increases, i.e.,
{	\begin{align}\label{equation49}
	\text{Pr}\left\{ P^{\text{AoI}}>\frac{A_{\text{th}}}{n}\right\}\approx e^{-\frac{A_{\text{th}}}{n}\theta_{\text{AoI}}}
\end{align}}
where $\theta_{\text{AoI}}>0$ is the peak-AoI bounded QoS exponent, which quantifies the rate of exponential decay of the violation probability in terms of the peak AoI violation probability within the framework of our proposed modeling schemes when considering the end-to-end  QoS provisionings including queuing delay, transmission delay, and propagation delays by implementing HARQ. 
Note} that the peak-AoI bounded QoS exponent is an indicator of the level of stringency associated with statistical peak-AoI bounded QoS provisioning as the threshold increases, which describes the relationship between the peak AoI threshold and the probability of peak AoI exceeding that threshold.

For a given designated peak AoI threshold, the peak AoI violation probability characterizes tail behaviors for the peak AoI, which can be used to compare with the delay and error-rate violation probabilities.
Accordingly, {a sufficiently small peak-AoI bounded QoS exponent enables the satellite-terrestrial integrated system to accommodate an arbitrarily large peak AoI, whereas a sufficiently large peak-AoI bounded QoS exponent renders the system intolerant of any peak AoI.}
Towards this end, to guarantee stringent mURLLC requirements, it is essential to consider the peak AoI violation probability as a new dimension of the QoS metric, which enables the analysis, management, and assessment of the data freshness {and alleviate the tail behaviors of the peak AoI}.

\subsection{The Error-Rate Bounded QoS Metrics Using HARQ in the Finite Blocklength Regime}

In analyzing the reliability issues, we investigate the statistical QoS requirement that describes the tail behavior of the probability of the decoding error with an increasing blocklength over satellite-terrestrial integrated networks when using HARQ.
In the realm of small-packet communications, the tradeoff between reliability and throughput has conventionally been studied by characterizing the \textit{error-rate bounded QoS exponent}.
Consequently, it is imperative to investigate the correlations between the \textit{error-rate bounded QoS exponent} and the decoding error probability as a means of characterizing error-rate bounded QoS metric over satellite-terrestrial integrated wireless networks.

\textit{Definition 3: The error-rate bounded QoS exponent:} Based on the LDP, the \textit{error-rate bounded QoS exponent}, denoted by $\theta_{\text{error}}$, quantifies the rate of exponential decay of the QoS violation probability in terms of the decoding error probability $\epsilon(\mu)$, within the framework of our proposed modeling schemes for satellite-terrestrial integration, which is defined as follows:
\begin{equation}\label{equation01}
	\theta_{\text{error}}\triangleq\lim\limits_{\widehat{n}\rightarrow \infty}-\frac{1}{\widehat{n}}\log(\epsilon(\mu)).
\end{equation}

Our proposal centers on the modeling and assessment of fundamental performance, with a particular emphasis on various facets related to the error-rate bounded QoS exponent. This exponent delineates the rate at which the decoding error probability exponentially decays as the codeword blocklength tends to infinity, {i.e.,
\begin{equation}\label{equation039}
	\epsilon(u)\leq\exp(-n\theta_{\text{error}}).
\end{equation}
The error-rate-bounded QoS exponent exhibits a consistent trend of decrement as the blocklength increases.} 
 The error-rate bounded QoS exponent plays a critical role in establishing the \textit{fundamental performance boundaries} for statistical QoS provisioning.
{In addition, the error-rate-bounded QoS exponent is determined by the channel capacity, the coding rate, and the channel dispersion.}

However, how to analytically derive the closed-form expression of the error-rate bounded QoS exponent over satellite-terrestrial integrated wireless networks remains an open and challenging problem.
One potential solution is to apply the LDP and Laplace transform for characterizing the asymptotic behaviors of remote tails regarding the decoding error probability.

Correspondingly, by using FBC, there exists a \textit{tradeoff} between the peak AoI violation probability and the delay violation probability. Specifically, the Mellin transform of the service time intends to decrease monotonically, while the Mellin transform of the service process aims to increase monotonically as the decoding error probability function approaches zero.

\section{Performance Evaluations}\label{sec:results}

\begin{figure}[!t]
	\centering
		\captionsetup{labelformat=default,labelsep=space} 
	\includegraphics[scale=0.43]{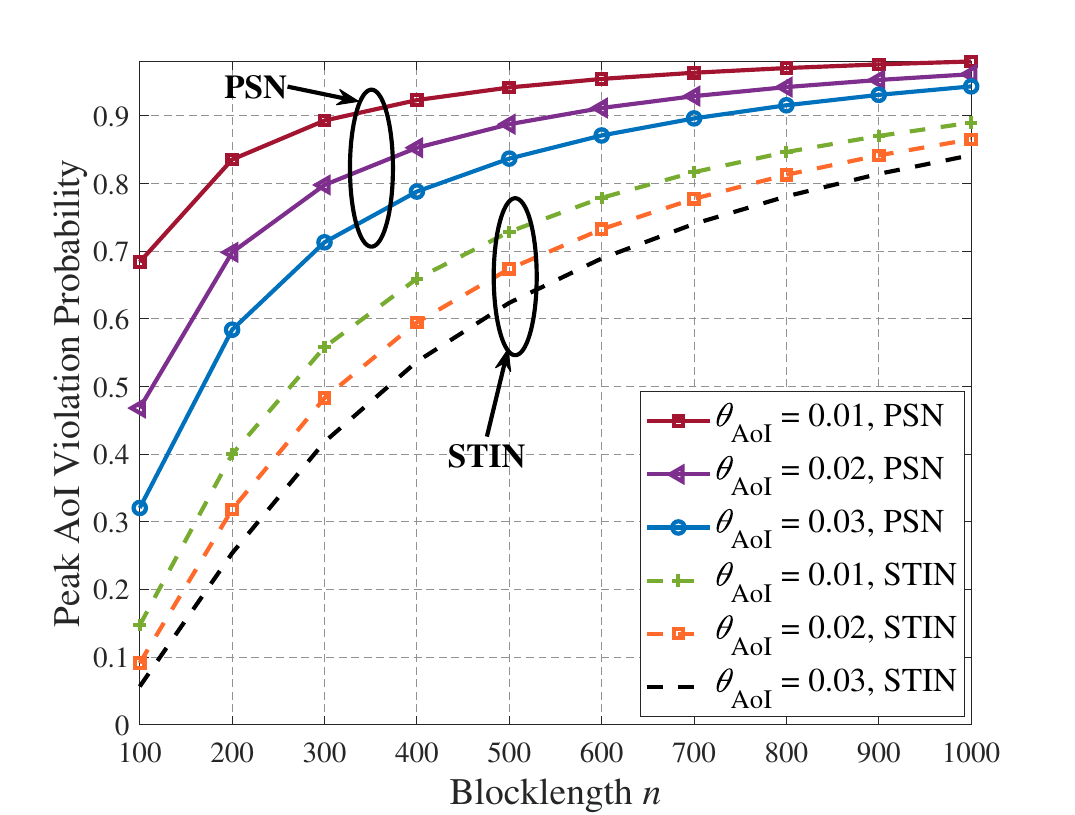}
	\caption{The peak AoI vs. the number of GBSs for our proposed satellite-terrestrial integrated networks.}
	\label{fig:03}
\end{figure}

We present a series of numerical results to verify and assess the effectiveness of our proposed statistical QoS provisioning schemes over satellite-terrestrial integrated 6G mobile wireless networks.
{Before the simulation, it is necessary to determine the relationship between channel use (cu) and second. 
Assuming that 2PSK modulation is employed and the transmission rate of the satellites is 1 Mbit/s, then we obtain 1 cu = $10^{-6}$ s. 
We assume that the nearby GBSs are located within a finite annular area centered around the destination node with an inner radius, denoted by $R_{\text{in}}$, and an outer radius, denoted by $R_{\text{out}}$, beyond which the interference is assumed to be negligible due to pathloss.
Throughout our simulations, we set the outer radius $R_{\text{out}}=10$ Km, inner radius $R_{\text{in}}=2$ km,  the antenna gain at satellite to be $20$ dBi, and the transmit power at the satellite$\in[10,50]$ dBm.}

Fig.~\ref{fig:03} depicts the peak AoI violation probability against the blocklength for the developed satellite-terrestrial integrated networks (STIN) as compared with the pure satellite networks (PSN).
Fig.~\ref{fig:03} indicates that the peak AoI violation probability exhibits an increasing trend concerning the blocklength.
In addition, Fig.~\ref{fig:03} shows that the proposed STIN achieves better peak AoI performance when compared to the PSN since the peak AoI in terrestrial systems is relatively lower than that of the PSN.
Fig.~\ref{fig:03} also illustrates that as the peak-AoI bounded QoS exponent $\theta_{\text{AoI}}$ increases, there is a corresponding reduction in the peak AoI.
This indicates that a smaller peak-AoI bounded QoS exponent $\theta_{\text{AoI}}$ establishes an upper bound, while a large peak-AoI bounded QoS exponent sets a lower bound on the peak AoI violation probability.

Fig.~\ref{fig:05} depicts the error-rate bounded QoS exponent against the blocklength for the developed performance modeling formulations.
The graph in Fig.~\ref{fig:05} demonstrates that as the blocklength approaches infinity, the error-rate bounded QoS exponent also decays exponentially, providing further confirmation of the definition of the error-rate bounded QoS exponent $\theta_{\text{error}}$.
Fig.~\ref{fig:05} also illustrates that when setting a large number of HARQ retransmissions $L$, there is an observable increase in the error-rate bounded QoS exponent. 
This suggests that the decoding error probability experiences a rapid decline as $L$ increases, indicating an augmented level of reliability.

\begin{figure}[!t]
	\centering 
		\captionsetup{labelformat=default,labelsep=space} 
	\includegraphics[scale=0.42]{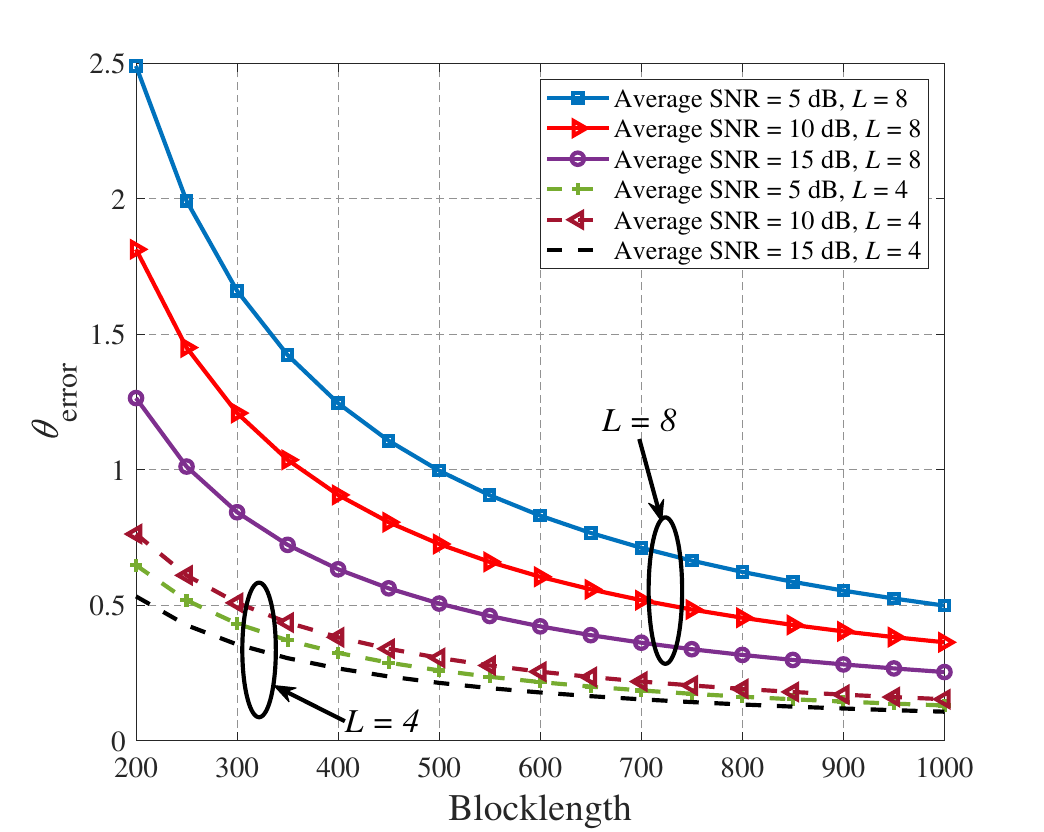}
	\caption{The error-rate bounded QoS exponent $\theta_{\text{error}}$ vs. the blocklength.}
	\label{fig:05}
\end{figure}

\section{Conclusions}\label{sec:conclusion}
We have proposed satellite-terrestrial integrated 6G mobile wireless networks architectures and developed a set of fundamental QoS metrics and their controlling techniques for statistical QoS provisioning in the finite blocklength regime.
Firstly, we have reviewed key techniques and potential solutions within the purview of 3GPP considering HARQ and AMC.
Secondly, using the HARQ protocol, a collection of novel fundamental statistical-QoS metrics, including the peak-AoI bounded QoS exponent, delay-bounded QoS exponent, and error-rate bounded QoS exponent are developed within the framework of FBC.
Finally, a series of simulations have been conducted to verify, assess, and examine our established statistical QoS provisioning schemes for satellite-terrestrial integrated 6G mobile wireless networks.

Conflict of interest statement. None declared.

\begin{IEEEbiography}[{\includegraphics[width=0.9in,height=1.25in,clip,keepaspectratio]{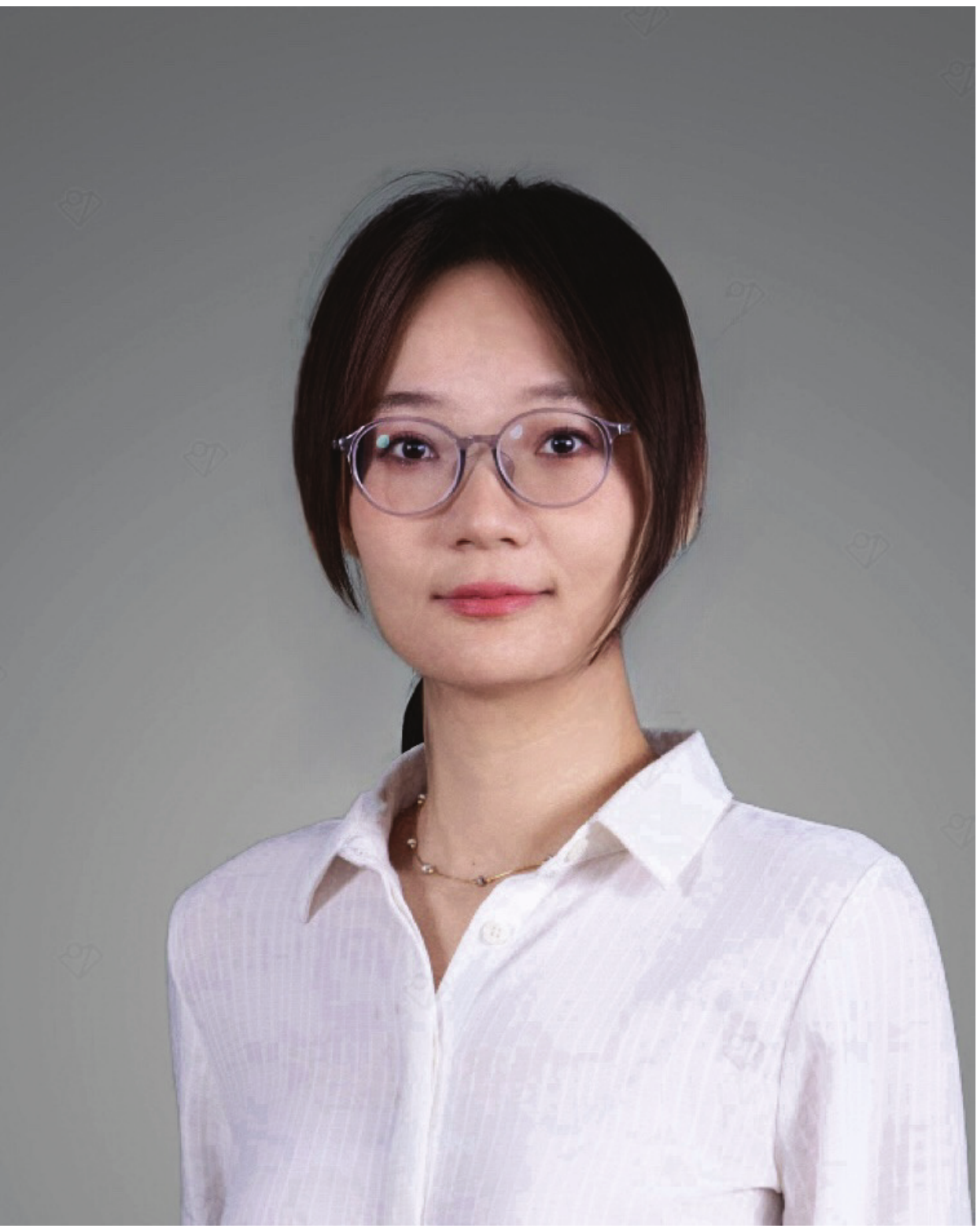}}]{Jingqing Wang} (Member, IEEE)  received the B.S. degree from Northwestern Polytechnical University, Xi'an, China, in Electronics and Information Engineering and the Ph.D. degree from Texas A\&M University, College Station, TX, USA, in Computer Engineering in 2022. She is currently a Jingying assistant professor with Xidian University. She has published more than 60 international journal and conference papers in IEEE JOURNAL ON SELECTED AREAS IN COMMUNICATIONS, IEEE magazines, IEEE TRANSACTIONS, IEEE INFOCOM, GLOBECOM, WCNC, and ICC. She won the Best Paper Award from the IEEE GLOBECOM in 2020 and 2014, respectively. Her current research interests focus on next generation mobile wireless network technologies, statistical QoS provisioning, 6G URLLC, information-theoretic analyses of FBC, emerging machine learning techniques.
\end{IEEEbiography}

\begin{IEEEbiography}[{\includegraphics[width=1in,height=1.28in,clip,keepaspectratio]{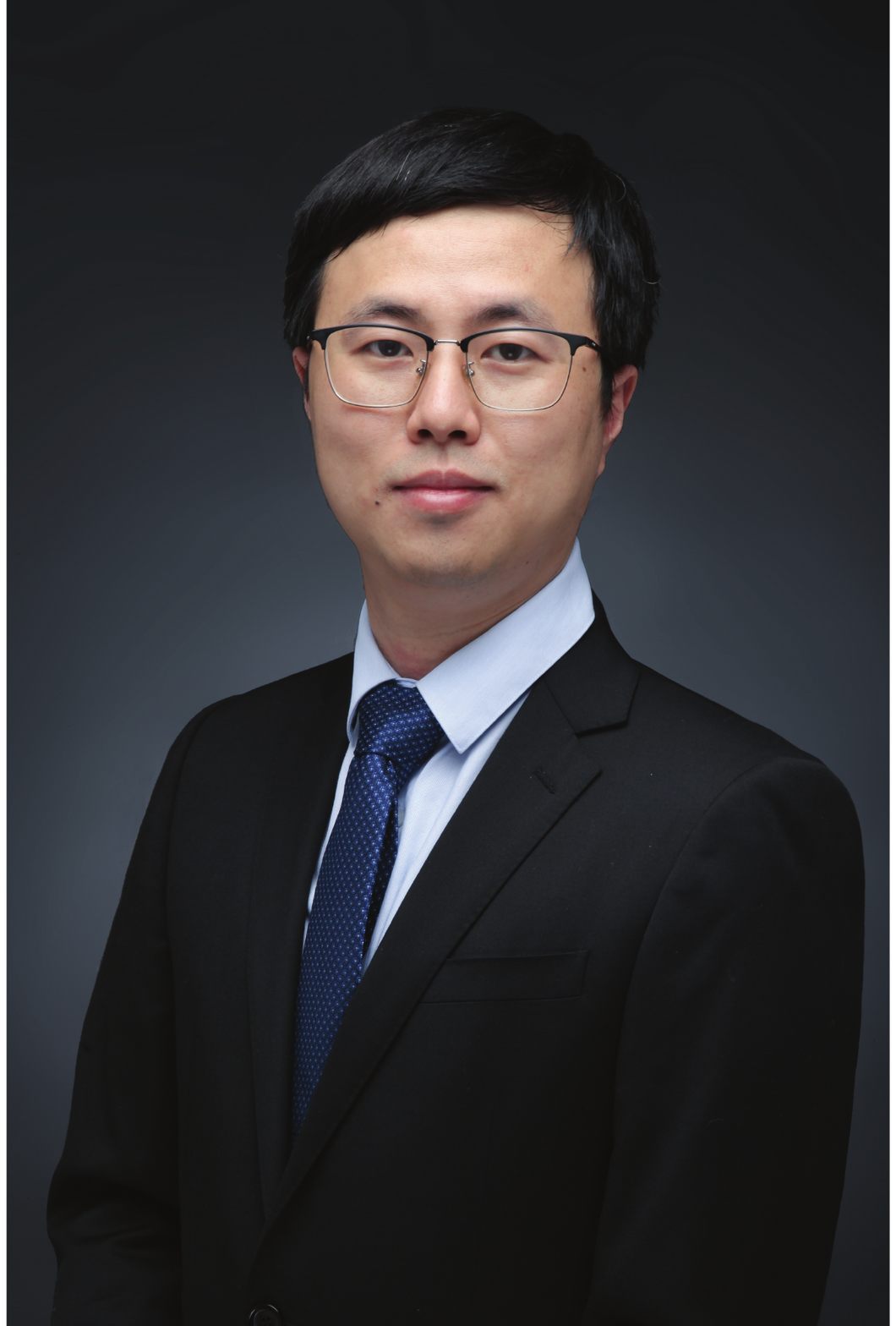}}]{Wenchi Cheng} (Senior Member, IEEE) received the B.S. and Ph.D. degrees in telecommunication engineering from Xidian University, Xian, China, in 2008 and 2013, respectively. He was a Visiting Scholar with the Department of Electrical and Computer Engineering, Texas A\&M University, College Station, TX, USA, from 2010 to 2011. He is currently a Full Professor with Xidian University. His current research interests include B5G/6G wireless networks, emergency wireless communications, and orbital-angular-momentum-based wireless communications. He has published more than 100 international journal and conference papers in IEEE JOURNAL ON SELECTED AREAS IN COMMUNICATIONS, IEEE magazines, IEEE TRANSACTIONS, IEEE INFOCOM, GLOBECOM, and ICC. He received the IEEE ComSoc Asia-Pacific Outstanding Young Researcher Award in 2021, the URSI Young Scientist Award in 2019, the Young Elite Scientist Award of CAST, and four IEEE journal/conference best papers. He has served or serving as the Wireless Communications Symposium Co-Chair for IEEE ICC 2022 and IEEE GLOBECOM 2020, the Publicity Chair for IEEE ICC 2019, the Next Generation Networks Symposium Chair for IEEE ICCC 2019, and the Workshop Chair for IEEE ICC 2019/IEEE GLOBECOM 2019/INFOCOM 2020 Workshop on Intelligent Wireless Emergency Communications Networks. He has served or serving as an Associate Editor for IEEE SYSTEMS JOURNAL, IEEE COMMUNICATIONS LETTERS, and IEEE WIRELESS COMMUNICATIONS LETTERS.
\end{IEEEbiography}

\begin{IEEEbiography}[{\includegraphics[width=1in,height=1.28in,clip,keepaspectratio]{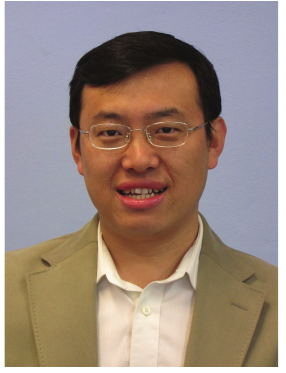}}]{Wei Zhang} (Fellow, IEEE) received the Ph.D. degree from The Chinese University of Hong Kong in 2005. He is currently a Professor at the School of Electrical Engineering and Telecommunications, University of New South Wales, Sydney, Australia. His current research interests include UAV communications, 5G, and beyond. He has served as a member in various ComSoc boards/standing committees, including Journals Board, Technical Committee Recertification Committee, Finance Standing Committee, Information Technology Committee, and Steering Committee for IEEE TRANSACTIONS ON GREEN COMMUNICATIONS AND NETWORKING and IEEE NETWORKING LETTERS. He received six best paper awards from the IEEE conferences and ComSoc technical committees. He is currently serving as an Area Editor for the IEEE TRANSACTIONS ON WIRELESS COMMUNICATIONS and the Editor-in-Chief of Journal of Communications and Information Networks. Previously, he has served as an Editor for IEEE TRANSACTIONS ON COMMUNICATIONS, IEEE TRANSACTIONS ON WIRELESS COMMUNICATIONS, IEEE TRANSACTIONS ON COGNITIVE COMMUNICATIONS AND NETWORKING, and IEEE JOURNAL ON SELECTED AREAS IN COMMUNICATIONS COGNITIVE RADIO SERIES. Within the IEEE ComSoc, he has taken many leadership positions, including the Member-at-Large on the Board of Governors from 2018 to 2020, the Chair of the Wireless Communications Technical Committee from 2019 to 2020, the Vice Director of the Asia Pacific Board from 2016 to 2021, the Editor-in-Chief of the IEEE WIRELESS COMMUNICATIONS LETTERS from 2016 to 2019, the Technical Program Committee Chair of APCC 2017 and ICCC 2019, and the Award Committee Chair of the Asia Pacific Board and the Technical Committee on Cognitive Networks. He was an IEEE ComSoc Distinguished Lecturer from 2016 to 2017. He is the Vice President of the IEEE Communications Society.
	
\end{IEEEbiography}

\begin{thebibliography}{1}
	{		\bibitem{9861699}
AZARI M M, SOLANKI D, CHATZINOTAS, et al. Evolution of non-terrestrial networks from 5G to 6G: a survey[J]. IEEE Communications Survey \& Tutorials, 2022, 24(4): 2633-2672.

	
\bibitem{9311792}	
HAN	H, JIANG X, LU W, et al. A multi-agent reinforcement learning approach for massive access in NOMA-URLLC networks[J]. IEEE Transactions on Vehicular Technology, 2023, 72(12): 16 799-16 804.}
	
	
	\bibitem{9834874}
ZHANG X, WANG J, POOR H V. Statistical Delay and Error-Rate Bounded {QoS} Control for {URLLC} in the Non-Asymptotic Regime[C]//Proceedings of IEEE International Symposium on Information Theory (ISIT). Piscataway: IEEE Press, 2022: 2112-2117.

	
	
		{	\bibitem{Age2022}
FANG Z, WANG J, REN Y, et al. Age of Information in Energy Harvesting Aided Massive Multiple Access Networks[J]. {IEEE Journal on Selected Areas in Communications}, 2022, 40(5): 1441-1456.
		

	\bibitem{9109636}
LIAO H, ZHOU Z; JIA Z, et al. Ultra-Low AoI Digital Twin-Assisted Resource Allocation for Multi-Mode Power IoT in Distribution Grid Energy Management[J]. {IEEE Journal on Selected Areas in Communications}, 2023, 41(10): 3122-3132.}

	
	{	\bibitem{Poor2023}
GAO	J, WU Y, SHAO S, et al. Energy Efficiency of Massive Random Access in MIMO Quasi-Static Rayleigh Fading Channels With Finite Blocklength[J]. {IEEE Transactions on Information Theory}, 2023, 69(3): 1618-1657.

	\bibitem{2023performance}
YUAN L, DU Q, YANG N, et al. Performance Analysis of IRS-Aided Short-Packet NOMA Systems Over Nakagami-$m$ Fading Channels[J]. {IEEE Transactions on Vehicular Technology}, 2023, 72(6): 8228-8233.}

	
	\bibitem{9686609}
LI D, WU S, JIAO J, et al. Age-oriented transmission protocol design in space-air-ground integrated networks[J]. {IEEE Transactions on Wireless Communications}, 2022, 21(7): 5573-5585.
 	
	{	 	\bibitem{Adaptive2022}
 CHENG W, XIAO Y, ZHANG S, et al. Adaptive Finite Blocklength for Ultra-Low Latency in Wireless Communications[J]. {IEEE Transactions on Wireless Communications}, 2022, 21(6): 4450-4463.
 
 	\bibitem{5GNR}
3GPP. Physical channels and modulation (v16.6.0): TS 38.211[S]. 2021.}





\end{thebibliography}
\end{document}